\documentclass[aps,prl,twocolumn,superscriptaddress,groupedaddress]{revtex4}  

\usepackage{graphicx}  
\usepackage{dcolumn}   
\usepackage{bm}        
\usepackage{amssymb}   
\usepackage{epstopdf}
\usepackage{subfig}
\usepackage{verbatim}
\usepackage{amsmath}
\usepackage{xspace}
\usepackage{xcolor}

\def\ee{\ensuremath{e^+e^-}\xspace}
\def\qq{\ensuremath{q\bar{{q}}}\xspace}
\def\WW{\ensuremath{W^+W^-}\xspace}

\def\ntrkoff{\ensuremath{{\rm N}_{\rm Trk}^{\rm Offline}}\xspace}
\def\ntrk{\ensuremath{{\rm N}_{\rm trk}}\xspace}

\def\dphi{\ensuremath{\Delta\phi}\xspace}

\begin{document}

\widetext
\leftline{Version 10 as of \today}
\leftline{Accepted by PLB}

\newcommand{\pt}{$p_{\text{T}}$\xspace}

\title{Long-range near-side correlation in \ee Collisions at 183-209 GeV with ALEPH Archived Data}

\author{Yu-Chen Chen}
\affiliation{Massachusetts Institute of Technology, Cambridge, Massachusetts, USA}%

\author{Yi Chen}
\affiliation{Massachusetts Institute of Technology, Cambridge, Massachusetts, USA}%

\author{Anthony Badea}
\affiliation{University of Chicago, Chicago, Illinois, USA}%

\author{Austin Baty}
\affiliation{University of Illinois Chicago, Illinois, USA}%


\author{Gian Michele Innocenti}
\affiliation{Massachusetts Institute of Technology, Cambridge, Massachusetts, USA}%

\author{Marcello Maggi}
\affiliation{INFN Sezione di Bari, Bari, Italy}%

\author{Christopher McGinn}
\affiliation{Massachusetts Institute of Technology, Cambridge, Massachusetts, USA}%

\author{Michael Peters}
\affiliation{Massachusetts Institute of Technology, Cambridge, Massachusetts, USA}%

\author{Tzu-An Sheng}
\affiliation{Massachusetts Institute of Technology, Cambridge, Massachusetts, USA}%

\author{Jesse Thaler}
\affiliation{Massachusetts Institute of Technology, Cambridge, Massachusetts, USA}%

\author{Yen-Jie Lee}
\email{yenjie@mit.edu}
\affiliation{Massachusetts Institute of Technology, Cambridge, Massachusetts, USA}%

\date{\today}

\begin{abstract}
The first measurement of two-particle angular correlations for charged particles with LEP-II data is presented. The study is performed using archived hadronic \(e^+e^-\) data collected by ALEPH at center-of-mass energies up to 209 GeV, above the \WW production threshold, which provide access to unprecedented charged-particle multiplicities and more complex color-string configurations if compared to previous measurements at LEP-I energies. An intriguing long-range near-side excess is observed in the correlation function measured with respect to the thrust axis in the highest multiplicity interval (\(\rm N_{\mathrm{trk}}\geq 50\)). Such a structure is not predicted by the Monte-Carlo simulation. The harmonic anisotropy coefficients \(v_n\), which result from the Fourier expansion of the two-particle correlation functions, were also measured for the first time in \(e^+e^-\) data, and compared to {\sc pythia}~6 predictions and to the results obtained in proton-proton collisions. The results presented in the Letter provide novel experimental constraints on the formation of collective phenomena in point-like \(e^+e^-\) collisions.

\end{abstract}

\maketitle


In heavy-ion collision experiments, two-particle angular correlations~\cite{STAR:2005ryu,STAR:2009ngv,PHOBOS:2009sau,Chatrchyan:2012wg,Aamodt:2011by,Adam:2019woz} are extracted for studying the Quark-Gluon Plasma (QGP)~\cite{Busza:2018rrf}. In these measurements, a long-range angular correlation, known as the ridge~\cite{STAR:2009ngv,PHOBOS:2009sau}, has been observed in various collision systems and at different collision energies. Since the beginning of LHC operations, this ridge structure has also been observed in high-multiplicity proton-proton (pp) collisions by the CMS collaboration~\cite{Khachatryan:2010gv} and confirmed by other experiments at the LHC and RHIC using smaller collision systems than ion-ion collisions, such as pp~\cite{Aad:2015gqa}, proton-ion (pA)~\cite{CMS:2012qk,ALICE:2012eyl,ALICE:2013snk,ATLAS:2012cix,Aaij:2015qcq}, and deuteron-ion~\cite{PHENIX:2013ktj,STAR:2015kak,PHENIX:2018lia} collisions. In head-on heavy-ion collisions, the ridge structure is associated with the fluctuating initial state of the ions~\cite{Ollitrault:1992bk, Alver:2010gr}. However, the physical origin of the ridge structure in small systems remains under debate~\cite{Dumitru:2010iy,Dusling:2013qoz,Bozek:2011if,He:2015hfa,Nagle:2018nvi}. The potential correlations in the initial state partons arising from hadronic structure make understanding pp and pA measurements challenging. Numerous theoretical models exist to explain these systems with high particle densities. These models incorporate various mechanisms, from initial state correlations as suggested in~\cite{Dusling:2013qoz}, through final-state interactions~\cite{He:2015hfa}, to hydrodynamic effects~\cite{Bozek:2011if}. 

Lately, the focus has intensified on assessing two-particle correlations in even smaller systems than pp and pA collisions. This includes systems like photonuclear collisions with ultra-peripheral proton-lead and lead-lead data as demonstrated by ATLAS and CMS~\cite{ATLAS:2021jhn,CMS:2022doq}, electron-proton collisions reported by ZEUS~\cite{ZEUS:2019jya}, and \(e^+e^-\)~\cite{Badea:2019vey,Belle:2022fvl,Belle:2022ars}. Such studies are invaluable complements to those done on larger collision systems, shedding light on the bare minimum conditions required for collective behavior~\cite{Nagle:2017sjv}. Electron beams, in particular, are free from issues like multiple parton interactions and initial state correlations. Notably, no significant ridge-like patterns have been detected so far in electron-positron annihilations, giving further experimental constraints to the emergence of the collectivity signal, as discussed in various theoretical works
~\cite{Bierlich:2019wld,Bierlich:2020naj,Castorina:2020iia,Agostini:2021xca,Larkoski:2021hee,Baty:2021ugw}. 

There are two potential approaches to making progress in detecting a possible ridge-like signal. The first approach involves increasing the final state multiplicity of the system. This is because the probability of parton-parton (or hadron-hadron) scattering increases with rising parton (hadron) density and larger final state multiplicity. Insights gained from pp and photonuclear collisions also suggest that a larger multiplicity decreases the magnitude of the negative direct flow ($v_1$) due to momentum conservation~\cite{ATLAS:2021jhn,CMS:2022doq}. A diminished $v_1$ can facilitate the detection of the possible ridge-like signal. The second approach involves testing the emergence of collective effects in the presence of different physics processes and more complex color-string configurations. As suggested in Ref.~\cite{Nagle:2017sjv}, a two-string configuration simulated in {\it A Multi-Phase Transport (AMPT)} framework strengthens the ridge-like signal compared to a single-string configuration. Investigating data with a two-string configuration could increase the chances of detecting a ridge-like signal in the most elementary collisions. The use of the LEP-II data with collision energy up to \(\sqrt{s} = 209\) GeV offers an obvious advantage in this exploration. Above the \WW production threshold (\(\sqrt{s} = 160\)~GeV), four-fermion processes mediated by either single or double $W$ or $Z$ bosons serve as subdominant channels in hadronic decays. This provides more complexity than the dominant single-string \(e^{+}e^{-} \to \gamma^{*}/Z \to \qq\) configuration. Particularly at the highest multiplicity, \(\WW \to 4q\) production emerges as the dominant channel.

This study utilizes archived data collected by the ALEPH detector at LEP-II~\cite{Decamp:1990jra} between 1996 and 2000. To analyze these data, an MIT Open Data format was created~\cite{Tripathee:2017ybi}. Unlike the 91.2 GeV sample at LEP-I~\cite{Badea:2019vey,Chen:2021uws}, which is dominated by $Z$-decays, the LEP-II sample sees significant contributions from various processes beyond $e^+e^- \to q\bar{q}$ fragmentation, including a notable ``radiative-return-to-$Z$'' effect due to QED initial-state radiation (ISR). Adopting the selection criteria from the ALEPH collaboration~\cite{ALEPH:1999iel}, we cluster the event into two jets to determine the effective center-of-mass energy ($\sqrt{s'}$) using the equation 
\begin{equation}
s' = \frac{\sin \theta_1 + \sin \theta_2 - | \sin(\theta_1 + \theta_2) |}{ \sin \theta_1 + \sin \theta_2 + | \sin(\theta_1 + \theta_2) | } \times s,
\end{equation}
where $\theta_{1,2}$ are the angles of these jets to the beam direction. Using this, the visible two-jet invariant mass ($M_{\rm vis}$) is derived, aiding in minimizing the QED ISR background. In our analysis, $\sqrt{s'}$ must exceed $0.9 \sqrt{s}$, and $M_{\rm vis}$ must surpass $0.7 \sqrt{s}$. Furthermore, adhering to the hadronic event criteria from previous LEP-I work~\cite{Badea:2019vey, ALEPH:2003obs}, events are selected based on the event sphericity axis's polar angle ($7\pi/36<\theta_{\rm lab}< 29\pi/36$), and those with under five tracks or with total reconstructed charged-particle energy below $15$~GeV are discarded.

High-quality tracks are selected using requirements identical to those in previous ALEPH analyses~\cite{Barate:1996fi}. They are also required to have a transverse momentum with respect to the beam axis ($p_{\rm T}^{\rm lab}$) above 0.2~GeV/$c$ and $|\cos{\theta_{\text{lab}}}|<0.94$ in the lab frame. We employed the Monte Carlo (MC) events from the ALEPH collaboration for reconstruction effects and data correction. 
Monte Carlo events were simulated using dedicated generators to model different hard processes~\cite{Jadach:1999vf,Jadach:1995nk,Jadach:2001mp,Sjostrand:2000wi,ALEPH:1993gxq} and weighted according to their cross sections. The descriptions of the parton fragmentation and hadronization were performed using $\textsc{pythia}$~6.1~\cite{Sjostrand:2000wi} or $\textsc{jetset}$~7.4~\cite{Sjostrand:1993yb}.
The ALEPH MC simulation was shown to provide an excellent description of both QCD and electroweak observables~\cite{ALEPH:2004dmh,ALEPH:2006cdc,ALEPH:2010iye, ALEPH:2003obs}. 

 
The analysis procedure aligns with prior two-particle correlation function studies~\cite{CMS:2012qk, Badea:2019vey}. For each event, the efficiency-corrected differential yield of charged-particle pairs, denoted as $\frac{\rm d^2 N^{\rm same}}{\rm d\Delta\eta \rm d\Delta\phi}$ (where ``same'' means particles from the same event), is computed. It is then normalized by the average corrected number of charged particles in the event, $\rm N_{\text{trk}}^{corr}$, yielding:
\begin{align}
S(\Delta\eta,\Delta\phi) &= \frac{1}{\rm N_{\rm trk}^{corr}}\frac{\rm d^2 N^{\rm same}}{\rm d\Delta\eta \rm d\Delta\phi}.
\end{align}

\begin{figure}[t]%
    \centering
    \includegraphics[width=0.48\textwidth]{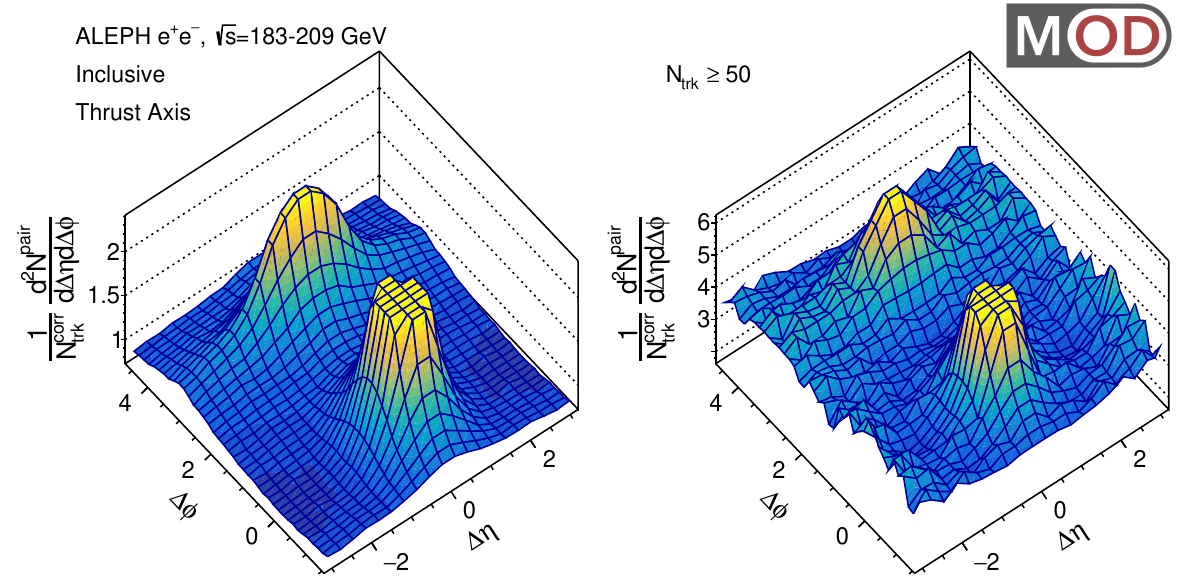}
    \caption{Two-particle correlation functions for events with the number of charged particle tracks in hadronic \ee in the thrust coordinate analysis with $\rm N_{\mathrm{trk}}\geq$ 5 (left) and $\rm N_{\mathrm{trk}}\geq$ 50 (right). The result is obtained considering pairs of tracks with transverse momentum with respect to the beam axis ($p_{\rm T}^{\rm lab}$) above 0.2~GeV/$c$. The sharp near-side peaks arise from jet correlations and have been truncated to illustrate the structure outside that region better.}%
    \label{fig:2PC_BeamThrust}%
\end{figure}
A mixed-event background correlation, $B(\Delta\eta,\Delta\phi)$, pairs charged particles from one event with those from 48 random events of the same multiplicity, giving
\begin{align}
B(\Delta\eta,\Delta\phi) &= \frac{1}{\rm N_{\rm trk}^{corr}}\frac{\rm d^2 N^{\rm mix}}{\rm d\Delta\eta \rm d\Delta\phi}.
\end{align}
Here, ${\rm N^{\rm mix}}$ is the efficiency-corrected pair count from the mixed event. By dividing this by $B(0,0)$, computed using pairs with $|\Delta\eta|<0.32$ and $|\Delta\phi|<\pi/20$, we obtain the detector's pair acceptance for uncorrelated particles. Hence, the acceptance-corrected pair yield is:
\begin{align}
\frac{1}{\rm N_{\rm trk}^{corr}}\frac{\rm d^2N^{pair}}{d\Delta\eta \rm d\Delta\phi} &= B(0,0) \times \frac{S(\Delta\eta, \Delta\phi)}{B(\Delta\eta, \Delta\phi)}.
\end{align}
For multiplicity-dependent analysis, events are grouped into five intervals based on reconstructed charged track count, $\rm N_\text{trk}$, with $p_{\rm T}^{\rm lab}>0.2$ GeV/c. 
Details, including multiplicity ranges and average track counts before and after correction, are in Table~\ref{fig:table2}.

\begin{table}[t]\centering
\begin{tabular}{ cccc}
 \hline
 $\rm N_{\text{trk}}$ range & Fraction of data (\%) & $\langle {\rm N_{\text{trk}}}\rangle$ & $\langle {\rm N_{\text{trk}}^{\text{corr}}}\rangle$\\
 \hline
 $\left[ \mathrm{10,20} \right)$&   58.6& 15.2& 17.3\\
 $\left[ \mathrm{20,30} \right)$&   33.1& 23.1& 25.7\\
 $\left[ \mathrm{30,40} \right)$&   3.7& 32.6& 35.9\\
 $\left[ \mathrm{40,50} \right)$&   0.4& 42.8& 47.1\\
 $\left[ \mathrm{50,\infty} \right)$&   $<0.1$& 53.0& 58.4\\
 \hline
\end{tabular}
\caption{Fraction of the full event sample for each multiplicity class. The last two columns
show the observed and corrected multiplicities, respectively, of charged particles with $p_{\rm T}^{\rm lab} >$ 0.2 GeV/c and $|\cos{\theta_{\text{lab}}}|<0.94$.}
\label{fig:table2}
\end{table}
Experimentally, the thrust axis~\cite{Farhi:1977sg} estimates the direction of the outgoing-state energy flow in \ee collisions, and it is used to define the coordinate system adopted by the thrust-axis analysis. The determination of the thrust-axis direction involves all particles within each event, along with the missing transverse energy (MET). By including the MET in the calculation, one can mitigate the effect of detector inefficiencies and improve the thrust resolution. All tracks meeting quality criteria then have their kinematic variables (\pt, $\eta$, $\phi$) recalculated, with the thrust axis substituting the beam axis, using the prescription of the LEP-I analysis~\cite{Badea:2019vey}. The variation of the thrust axis direction causes the ALEPH detector acceptance in the thrust coordinates to vary on an event-by-event basis. This is accounted for by recalculating the kinematics variables for particles in paired events with respect to the thrust axis in the signal event. The $\eta$ and $\phi$ distributions of the charged tracks in the paired events are then reweighted to match those of signal events. 

In hadronic collision systems, the azimuthal anisotropy of charged particle production is typically quantified with the harmonic anisotropy coefficients $v_{n}$~\cite{Voloshin:1994mz,Poskanzer:1998yz,Alver:2010gr}.  In particular, the second-order elliptic coefficient, $v_{2}$ is sensitive to the collective behavior and the level of thermalization of the system in relativistic heavy ion collisions~\cite{Ollitrault:1992bk,Ackermann:2000tr}. However, it is often difficult to make a direct quantitative connection between the size of any associated yields and the corresponding value of $v_{2}$ because most of the structure of the correlation functions comes from jet-like correlations.  These correlations are sometimes referred to as ``nonflow''~\cite{Adler:2002tq,Adare:2008ae,Aamodt:2010pa,Sirunyan:2017pan}.  

We employ the Fourier decomposition analysis used in prior studies to investigate potential flow-like signatures. This helps us constrain anisotropy harmonics through two-particle azimuthal correlations. The non-flow effects diminish significantly at large $|\Delta \eta|$. The long-range azimuthal differential yields can be described by:
\begin{align}
 Y_l(\Delta\phi) = \frac{1}{{\rm N}_{\rm trk}^{\rm corr}}\frac{d{\rm N}^{\rm pair}}{d\Delta\phi} = \frac{{\rm N}^{\rm assoc}}{2\pi} \left( 1 + \sum_{n=1}^{\infty} 2 V_{n\Delta} \cos(n\Delta\phi) \right), 
\label{eq:ydphi}
\end{align}
with ${\rm N}^{\rm assoc}$ representing associated track pairs in specified $|\Delta \eta|$ and $\Delta \phi$ ranges. The long-range associated yield is computed using a histogram, and the Discrete Fourier Transform is used to determine Fourier coefficients ($V_{n\Delta}$) and normalization (${\rm N}^{\rm assoc}$). 
The single-particle Fourier harmonics $v_n$ can be extracted as $v_n = \text{sign}(V_{n\Delta})\sqrt{|V_{n\Delta}|}$, assuming factorization~\cite{Voloshin:1994mz}.

This analysis uses Bayesian inference~\cite{Bayesian} to assess the statistical uncertainties for the observables of interest: correlation yields and harmonic anisotropy coefficients $v_n$. The primary rationale behind adopting the Bayesian analysis is to offer a more detailed estimation of uncertainties, particularly when assuming a Gaussian distribution is not ideal for a data set with a non-Gaussian distribution. With Bayes' theorem, we obtain the posterior probability for an observable of interest, using a flat prior and a ``weighted Poisson distribution~\cite{Bohm:2013gla}'' as the likelihood function. Reported central values and uncertainties for pairing yields and harmonic anisotropy coefficients are based on the ``maximum a posteriori (MAP)'' method.
The comprehensive Bayesian calculation has been documented in the note~\cite{Chen:2023nsi}.


Systematic uncertainties for the long-range associated yield \(Y_l(\Delta \phi)\) and $v_n$ arise from event and track selections, the $B(0,0)$ normalization factor, and residual MC corrections. 
The uncertainty associated to the ISR selection was estimated by varying the requirements on the visible two-jet invariant mass $M_{\rm vis}$ from $0.7\sqrt{s}$ to $0.65\sqrt{s}$, and the value of the effective center-of-mass energy \(\sqrt{s'}\) from $0.9\sqrt{s}$ to $0.87\sqrt{s}$, following the strategy described in~\cite{ALEPH:2003obs}. 
The strategy for the estimation of the other systematic uncertainties is consistent with the one adopted for the LEP-I analysis approach~\cite{Badea:2019vey}. The systematic uncertainty on the hadronic event selection was obtained by varying the minimal number of particles from 13 to 10 and the reconstructed charged-particle energy from 15 GeV to 10 GeV. 
The total systematics uncertainty on the thrust-axis correlation functions that account for the ISR selections and the hadronic event selections is found to be below $0.50\%$ across different multiplicity bins. Varying the minimal requirement on the number of track hits in the time projection chamber from 4 to 7 leads to a systematic uncertainty smaller than $2\%$. Including the $B(0,0)$ factor as the normalization choice also introduces a systematic uncertainty. We evaluate its impact based on the statistical uncertainty of the $B(0,0)$ normalization factor, ranging from $0\%$ to $0.50\%$ for different multiplicity bins. Generally, these systematic uncertainties affect \dphi bins uniformly. Lastly, the residual MC correction factor results in an uncorrelated uncertainty across \dphi bins ascertained through different fit attempts on this correction factor. Three function types are evaluated, with half of their maximum deviation deemed as the associated uncertainty. The maximum deviation for the lowest multiplicity bin ($10 \leq \rm N_{\mathrm{trk}} \leq 20$) is $1.2\%$, while it is smaller than $0.1\%$ for higher multiplicity bins.
The contributions from the different sources described above are summed up in quadrature to compute the total systematic uncertainty on $Y_l (\Delta \phi)$, and are propagated to the determination of $v_n$ systematics. The resulting systematic uncertainties for each interval of \ntrkoff are summarized in Table~\ref{tab:Systematics_thrust}.

\begin{table}[ht] 
\footnotesize
\begin{center}
\begin{tabular}{ccccc}
\hline
\ntrkoff & TPC hits  & Event selections & $B(0,0)$  & Residual MC corr. \\
\hline
$[10,20)$ & 1.09 & 0.39 & 0.44 & 1.17 \\
$[20,30)$ & 0.68 & 0.44 & 0.21 & 0.11 \\
$[30,40)$ & 0.65 & 0.05 & 0.12 & 0.10 \\
$[40,50)$ & 0.73 & 0.04 & 0.16 & 0.13 \\
$[50,\infty)$ & 1.60 & 0.50 & 0.27 & 0.02 \\
\hline
\end{tabular} 
\end{center}
\caption{Systematic uncertainties as a function of the offline multiplicity \ntrkoff. All values are reported as percentages of the long-range differential associated yield.}
\label{tab:Systematics_thrust}
\end{table}



In Fig.~\ref{fig:2PC_BeamThrust}, the two-particle correlation functions for inclusive (left panel) and high multiplicity events (right panel) are shown. The result is obtained considering pairs of tracks with transverse momentum with respect to the beam axis ($p_{\rm T}^{\rm lab}$) above 0.2~GeV/$c$. No significant ridge-like structure was observed in the multiplicity-integrated result ($\ntrk \ge 5$). In the highest multiplicity bin ($\ntrk \ge 50$), an intriguing U shape was revealed at the large $|\Delta\eta|$ and small $\Delta\phi$ phase space, which is studied further in the later sections. 

\begin{figure}[tb]
    \centering
    \includegraphics[width=0.49\textwidth]{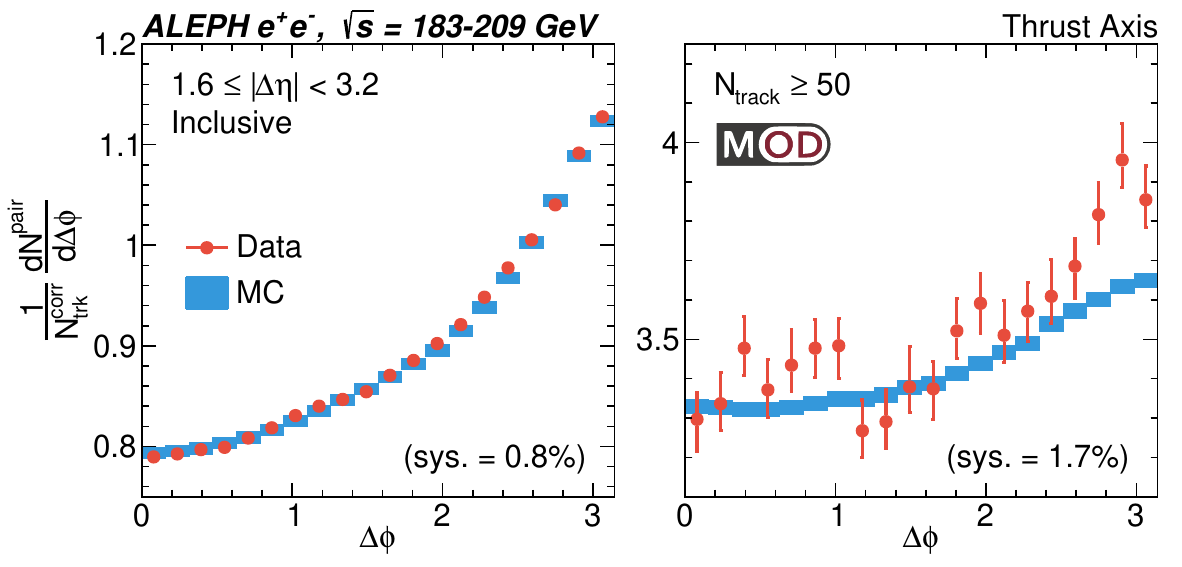}
    \caption{For the long-range region $1.6 < |\Delta\eta| < 3.2$, the azimuthal associated yield is presented for $N_{\rm trk} \ge 5$ (left) and $N_{\rm trk} \ge 50$ (right). Data is presented in red dots with statistical error bars, while systematic uncertainties are detailed in the text. The $\textsc{pythia}$~6 model is shown in blue with its statistical error band.}
\label{fig:dNdphi} 
\end{figure}

One-dimensional distributions in $\Delta\phi$ are studied by averaging the two-particle correlation function over the region between $1.6 < |\Delta\eta| < 3.2$ to investigate the long-range correlation in finer detail. Fig.~\ref{fig:dNdphi} shows the comparisons between data and MC on the long-range azimuthal differential associated yields for inclusive (left panel) and high-multiplicity events (right panel). The MC simulation describes well the measurement for $\ntrk \ge 5$. On the contrary, in the highest multiplicity interval ($\ntrk \ge 50$), the data reveals a long-range near-side structure that the MC simulation does not capture. Moreover, the data display a more significant slope when going to large $\dphi$ than predictions from MC.
We also examined the correlation functions using the $\textsc{pythia}$~8 simulation~\cite{Bierlich:2022pfr} with the default \textsc{monash 2013} tune~\cite{Skands:2014pea}. This framework allows for the inclusion of microscopic collective effects from the shoving mechanism~\cite{Bierlich:2017vhg,Bierlich:2020naj}. However, a similar long-range near-side enhancement is not seen in the $\textsc{pythia}$~8 simulations, either with or without the inclusion of the shoving model.

The long-range azimuthal differential distribution is then fitted within $0 < \Delta\phi < \pi/2$. Following the Zero Yield At Minimum (ZYAM) procedure, each distribution is then shifted down by the fit minimum ($c_{\text{ZYAM}}$)~\cite{Ajitanand:2005jj}. The associated yield is then quantified by integrating the subtracted distribution from $\Delta\phi = 0$ to the location of fit minimum $\Delta\phi_{min}$~\cite{Badea:2019vey,Chen:2023nsi}.

\begin{figure}[b!]
\begin{center}
\includegraphics[width=.4\textwidth]{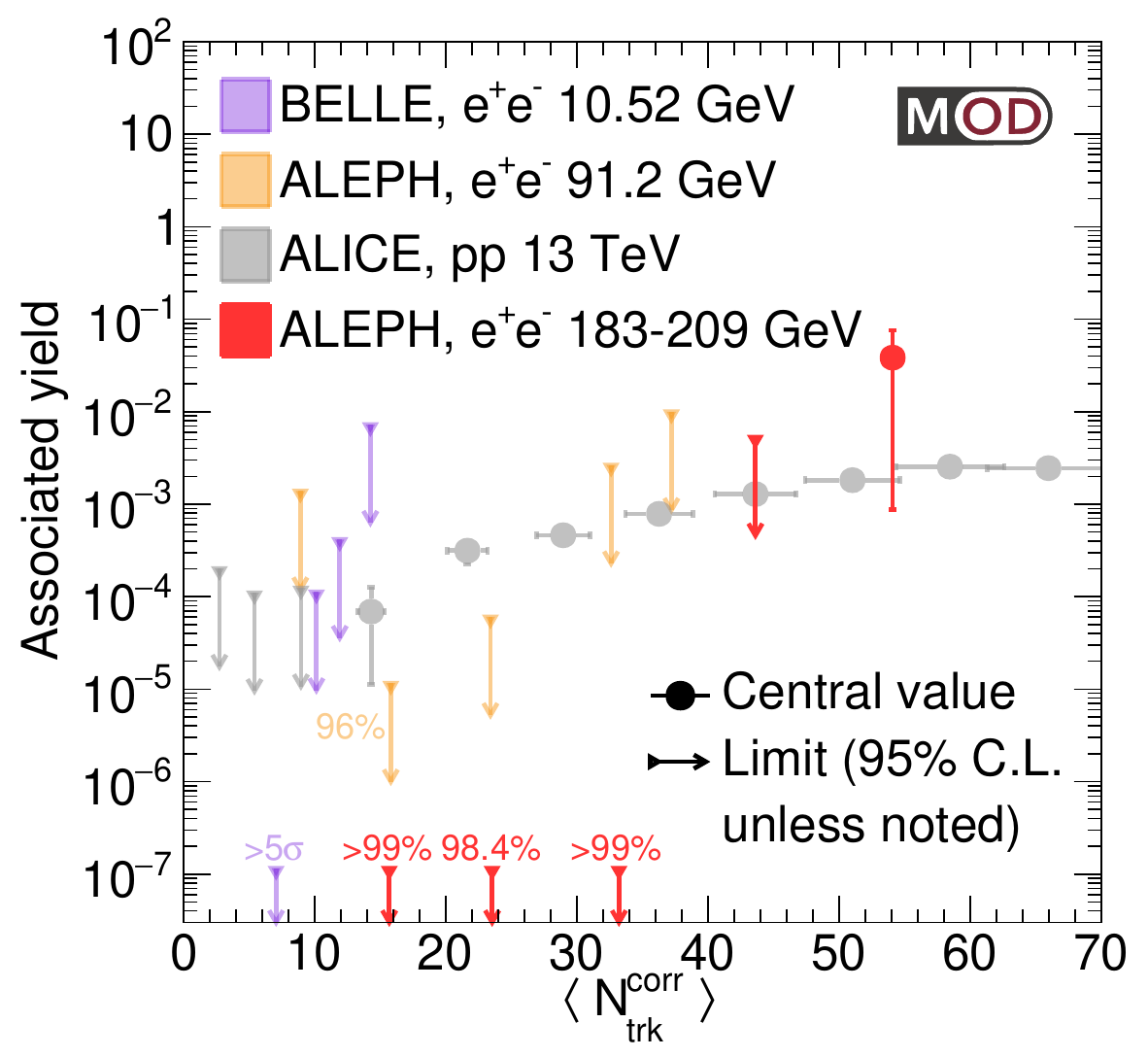}
\caption{Confidence limits on associated yield as a function of $\left \langle {\rm N}_{\mathrm{trk}}^{\mathrm{corr}} \right \rangle$ in the thrust axis analysis. This work (LEP-II analysis, $\sqrt{s}=183-209$~GeV) is shown in red, overlapping with results from Belle (pale purple)~\cite{Belle:2022fvl}, LEP-I (pale orange)~\cite{Badea:2019vey}, and ALICE (pale gray, lab frame)~\cite{ALICE:2023ulm}. 
The label ``$>5\sigma$'' indicates the $5 \sigma$ confidence level upper limit. The systematic uncertainties are included in the displays of confidence limits and the reported associated yield.}
\label{fig:confidenceLimits} 
\end{center}
\end{figure}

A bootstrap method~\cite{efron1979} was employed to estimate the uncertainties affecting the near-side long-range yields. This procedure generates variations in the $Y(\Delta\phi)$-distribution according to its statistical and systematic uncertainties. The uncorrelated systematic uncertainties include the event and track selection and the $B(0,0)$ variations, which mostly affect the normalization of the correlation function. The residual MC correction is treated in this procedure as a source of correlated systematic uncertainty across $\Delta\phi$ bins. A bootstrap sample of $2 \times 10^5$ variations was generated for each interval of ${\rm N}_{\rm trk}$. A near-side yield was extracted for each of these variations by exploiting a ZYAM fit method. This procedure was repeated considering three different fit functions, namely a three-term Fourier series plus a constant, a purely-even quartic function, and a purely-even quadratic function plus a $\cos{2\Delta\phi}$ term. The choice of fit function resulted as the dominant source of systematic uncertainty.

At low multiplicity, most variations lead to small associated yields. If over 5\% of variations exceed a yield of $1\times 10^{-7}$, we quote an upper limit at 95\% confidence level. Otherwise, a C.L. for variations below this threshold is stated. The aforementioned estimation is performed individually for the bootstrap samples generated with the three choices of fit functions. The most conservative confidence level (or confidence limit) is reported.

In the highest multiplicity interval, the central value is reported. The associated systematic uncertainty is obtained as the quadratic sum of the individual systematic sources. The measured associated yields in bins of multiplicity are shown in Fig.~\ref{fig:confidenceLimits}. The results are also overlaid with the measurements obtained in $e^{+}e^{-}$ collisions by Belle~\cite{Belle:2022fvl} and ALEPH (LEP-I)~\cite{Badea:2019vey}, and low-multiplicity pp collisions by ALICE~\cite{ALICE:2023ulm}. 
Incorporating the same scaling treatment for $e^{+}e^{-}$ and pp collisions as detailed in ALICE publication~\cite{ALICE:2023ulm}, we scale the $x$ axis of the ALICE data by the acceptance correction coefficients $c_{\rm ee} = 0.78$ and $c_{\rm pp} = 0.57$ for ALEPH and ALICE experiments, respectively. The scaled $\left \langle {\rm N}_{\mathrm{trk}}^{\mathrm{corr}} \right \rangle$ for ALICE data points are displayed with uncertainty ranges from the scaling process. The half of the maximum deviation between the correction coefficients is quoted as the relative uncertainty.
The reported thrust C.L.s are compatible or lower than the C.L.s and the central values of the associated yield reported by Belle, LEP-I, and ALICE. These C.L.s contrast measurements of a nonzero azimuthal anisotropy signal in lower multiplicity pp collisions~\cite{Aaboud:2016yar,CMS:2016fnw}. At a high multiplicity above 50, the results are compatible with pp results from ALICE.

\begin{figure}[t!]
\centering
    \includegraphics[width=0.49\textwidth]{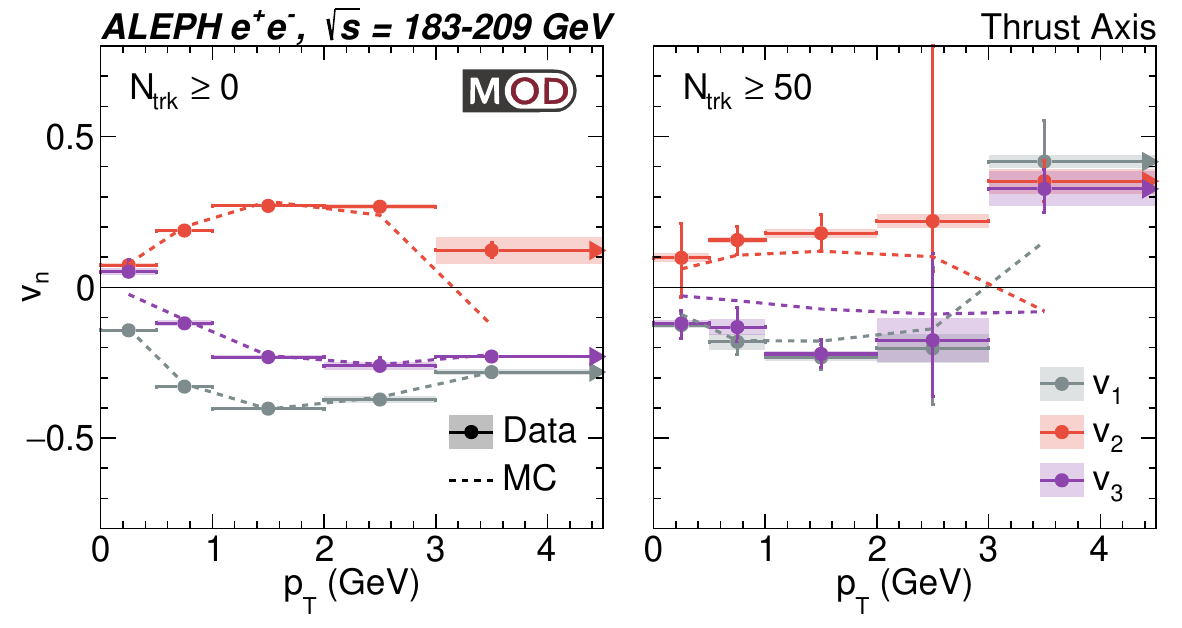}
\caption{$v_{n}$ as a function of the track pairs' $p_{T}$ requirement in different multiplicity intervals for the thrust axis analysis for the LEP-II high-energy sample. Data's $v_1$, $v_2$, and $v_3$ are shown in black, red, and purple error bars. MC results are dashed lines with corresponding colors.}
\label{fig:VnVsPt_thrust}
\end{figure}

\begin{figure}[htb]
\centering
    \includegraphics[width=0.4\textwidth]{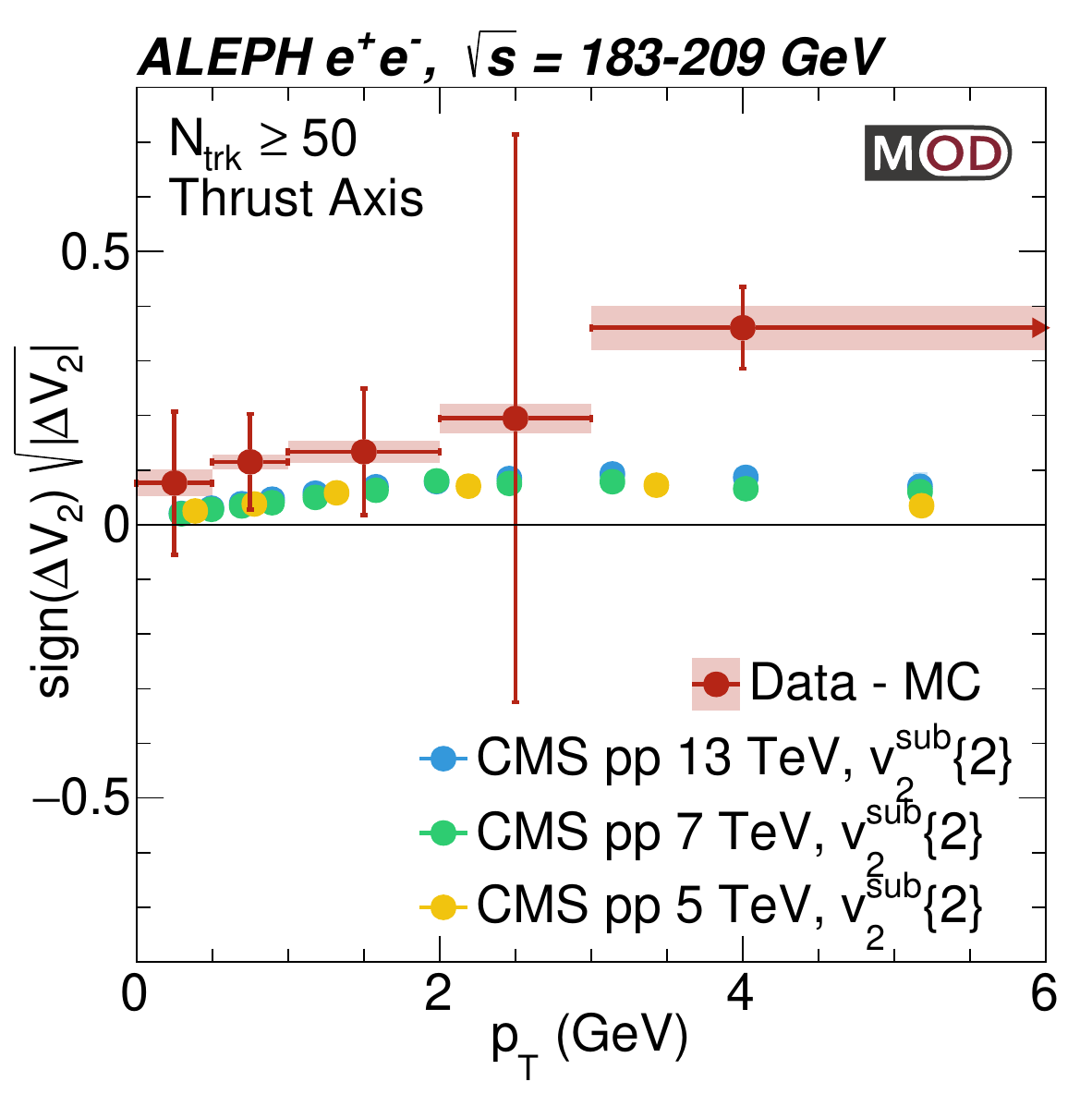}
\caption{Excess of elliptic anisotropy coefficient ${\rm sign} (\Delta V_2) \sqrt{\Delta V_2}$, where $\Delta V_2 = V_{2, \rm data} - V_{2, \rm MC}$, as a function of the track pairs' $p_{T}$ requirement for ${\rm N}_{\rm trk} \ge 50$ in the thrust axis analysis for LEP-II high-energy sample. The result is overlaid with CMS subtracted elliptic anisotropy coefficient measurements~\cite{CMS:2016fnw}.}
\label{fig:DeltaVn_overlay}
\end{figure}

In Fig.~\ref{fig:VnVsPt_thrust}, the $v_n$ coefficients as a function of $p_T$ are compared to the archived $\textsc{pythia}$~6 simulations. The multiplicity-integrated result (left panel), dominated by events with lower $\ntrk$, shows a decent agreement with the ALEPH MC simulation.
A difference with respect to the simulation is observed for events with $\ntrk \geq 50$, as shown in the right panel of the same figure. The simulation generally predicts a smaller magnitude for $|v_n|$. In Fig.~\ref{fig:DeltaVn_overlay}, the excess of harmonic anisotropy coefficient of data with respect to MC, defined as ${\rm sign} (\Delta V_2) \sqrt{\Delta V_2}$ where $V_2$ is a simplified notation of $V_{2\Delta}$ in Eq.~\ref{eq:ydphi}, is presented. This observable allows for suppressing jet-like correlations or any additional correlation that could emerge from known physics processes included in the $\textsc{pythia}$~6 simulations. Therefore, it provides enhanced sensitivity to new physics mechanisms, which are not modeled in the existing MC simulations and could induce changes in the elliptic anisotropy coefficient in data. The excess of harmonic anisotropy coefficient measured in \ee can also be used to obtain a qualitative comparison with the $v_2^{\text{sub}}\{2\}$ measured in pp collisions, where the effects of jet-like correlations are suppressed by using lower-multiplicity data or template-fit methods~\cite{CMS:2016fnw}.
The measured excess for events with $\ntrk \geq 50$ is shown in Fig.~\ref{fig:DeltaVn_overlay}, overlaid with the CMS high-multiplicity pp measurements of $v_2^{\text{sub}}\{2\}$ across three different collision energies~\cite{CMS:2016fnw}. Despite the qualitative nature of this comparison, the excess of elliptic anisotropy coefficient measured in \ee collisions and the $v_2^{\rm sub}\{2\}$ measured in pp collision shows, quite remarkably, a similar trend as a function of \pt.



In summary, we present the first measurement of two-particle angular correlations from \ee annihilation at energies $\sqrt{s} = 183$--$209$ GeV using archived ALEPH LEP-II data recorded between 1996 and 2000. In analyzing the thrust axis of these collisions between $\sqrt{s}=183$ to 209 GeV, a long-range near-side excess in the correlation function emerges. For the first time, we decomposed two-particle correlation functions in \ee collisions using a Fourier series. The resulting harmonic anisotropy coefficients $v_2$ and $v_3$ measured with LEP-II data with charged-hadron multiplicities above 50 are larger than those obtained from the archived MC simulation. We quantified this discrepancy with the excess harmonic anisotropy coefficient, defined as the difference between the data- and MC-derived $v_2$. The excess of harmonic anisotropy coefficient measured in high-multiplicity \ee collisions, measured as a function of track $p_T$, was compared to the $v_2^{\text{sub}}\{2\}$ measured in pp collisions~\cite{CMS:2016fnw}, where the contribution to the $v_2$ coming from jet-like correlations is subtracted using lower-multiplicity data. Despite the qualitative nature of the comparison and the limited statistical accuracy, the two measurements show a remarkably similar trend as a function of $p_T$ at high-multiplicity, where final-state interactions are expected to be involved in forming a ridge signal in hadronic collisions~\cite{ALICE:2021nir,Strickland:2018exs}. The work presented in the Letter motivates further theoretical and experimental efforts to identify the physics mechanisms that could be responsible for the emergence of such phenomena in \ee collisions. 
Future measurements performed in ultra-peripheral heavy-ion collisions and future experimental facilities, such as the Electron-Ion Collider at BNL or the Future Circular Collider at CERN, will also provide new and more differential constraints with increasing experimental accuracy to clarify the origin of long-range near-side correlations in small systems.

The authors would like to thank the ALEPH Collaboration for their support and foresight in archiving their data. We would like to thank the valuable comments and suggestions from Roberto Tenchini, Guenther Dissertori, Wei Li, Jiangyong Jia, Wit Busza, N\'estor Armesto, Jean-Yves Ollitrault, J\"urgen Schukraft and Jan Fiete Grosse-Oetringhaus. This work has been supported by the Department of Energy, Office of Science, under Grant No. DE-SC0011088 (to Y.-C.C., Y.C., M.P., T.S., C.M., Y.-J.L.), Eric and Wendy Schmidt AI in Science Postdoctoral Fellowship (A. Badea) and Grant No. DE-SC0012567 (to J.T.).

\bibliography{ridgepaperALEPH}

\end{document}